\magnification=1200
\baselineskip=20pt
\def\g{\gamma}
\def\b{\beta}
\def\a{\alpha}
\def\half{{1\over2}}
\def\bk{{\b^2\over k}}
\def\NI{\noindent}
\def\qa{{\hat q}_a}
\def\qb{{\hat q}_b}
\def\z{{\overline{z}}}
\def\k{{\overline{k}}}
\def\half{{1\over2}}
\def\kR{k_{\rm R}}
\def\kI{k_{\rm I}}
\def\C{{\rm\bf C}}
\def\O{{\cal O}}
\overfullrule = 0pt

\centerline{\bf The Solution of the Modified Helmholtz Equation in a Wedge}
\centerline{\bf and an Application to Diffusion-Limited Coalescence}
\bigskip
\centerline{Daniel ben-Avraham\footnote{$^1$}{{\bf e-mail:} qd00@clarkson.edu}}
\smallskip
\centerline{Physics Department, and Clarkson Institute for Statistical
Physics (CISP),} 
\centerline{Clarkson University, Potsdam, NY 13699-5820}
\medskip
\centerline{Athanassios S. Fokas\footnote{$^2$}{Permanent address:
Department of Mathematics, Imperial College, London SW7 2BZ,} \footnote{}{UK.
{\bf e-mail:} a.fokas@ic.ac.uk}}
\smallskip
\centerline{Institute for Nonlinear Studies,}
\centerline{Clarkson University, Potsdam, NY 13699-5805}

\bigskip\bigskip
\NI{\bf Abstract:}
The general solution of the modified Helmholtz
equation, $q_{xx}(x,y)+q_{yy}(x,y)-4\b^2q(x,y)=0$, in the wedge
$0\leq x\leq y\leq\infty$, is presented.  This solution is used to find the
explicit steady-state of diffusion-limited coalescence, $A+A\rightleftharpoons
A$, on the half-line, with a trap-source at the origin.
\smallskip\NI
02.60.L,\ 02.70,\ 61.43.H,\ 82.20.M

\vfill\eject

A new method for solving boundary value problems for linear and for
integrable nonlinear PDEs was introduced in~[1].  This method has been
applied so far to evolution equations~[2-4] and to Laplace's equation~[5].  In
this letter we apply this method to the equation
$$
E_{xx}+E_{yy}+\g(-E_x+E_y)=0\;,  \eqno(1)
$$
in the wedge $0\leq x\leq y$, where $E(x,y)$ is a scalar function and $\gamma$
is a positive constant.  Using the substitution
$E(x,y)=e^{-{\g\over2}(y-x)}q(x,y)+e^{-\gamma(y-x)}$, eq.~(1) becomes the
modified Helmholtz equation
$$
q_{xx}+q_{yy}-4\b^2q=0\;,\qquad 0\leq x\leq y\;, 
  \qquad \b={\gamma\over\sqrt{8}}\;.  \eqno(2)
$$
Eq.~(1) with $\gamma=v/D$ represents
the steady state of the diffusion-limited reaction $A+A\rightleftharpoons A$ on
the line, where the $A$-particles diffuse with diffusion constant $D$, they
merge immediately upon encounter, and split into two particles (the back
reaction) at rate $v$~[6-8].  $E(x,y)$ represents the
probability that the interval
$(x,y)$ is empty.  The concentration profile of the particles is related to
$E$ through $c(x)=-E_y(x,x)=\gamma-q_y(x,x)$.  Suppose that we limit ourselves
to  the positive $x$-axis, then the domain of Eq.~(1) is $0\leq x\leq y$.
The forward reaction is described by the boundary condition (BC) $E(x,x)=1$.
If the origin absorbs all impinging particles, but also 
generates particles at rate $v'$, then
$E_x(0,y)=\gamma'E(0,y)$, where $\gamma'=v'/D$.
The associated BCs for $q(x,y)$ are
$$
q(x,x)=0\;, \qquad
q_x(0,y)+({\gamma\over2}-\gamma')q(0,y)=(\gamma'-\gamma)e^{-{\gamma\over2}y}
   \;.   \eqno(3)
$$
Particular solutions are known for the case of a perfect trap at the
origin $(\gamma'=0)$, and for the case of $\gamma'=\gamma$. In the latter
case the steady state is identical to that of an infinite system, without a
trap. For $\gamma'>\gamma$, it is not clear {\it a priori\/}
whether a steady state even exists.  The present method shows that it does,
and it provides an explicit solution for all $\gamma\geq0$.

The implementation of the method of~[1] to eq.~(2) involves the following steps:
(a)~Construct a Lax pair for eq.~(2).  This means finding two linear
eigenvalue equations whose compatibility condition is eq.~(2).
(b)~Perform the {\it simultaneous\/} spectral analysis of this Lax pair.
This means constructing a function $\mu(x,y,k)$ which satisfies both equations
defining the Lax pair, and which is bounded for all complex values of the
spectral parameter $k$.  This can be used to: ($i$)~find an integral
representation of $q(x,y)$ in terms of certain functions $\qa(k)$ and $\qb(k)$,
which we refer to as spectral data; ($ii$)~find a {\it global relation\/}
satisfied by the boundary values of $q$.
(c)~Analyze the global relation.  This can be used to express $\qa(k)$
and $\qb(k)$ in terms of the given data on the boundary.

Steps~(a) and (b) are algorithmic.  The Lax pair, the integral
representation for $q(x,y)$, and the global relation, are given by eqs.~(6),
(14), (13).  We emphasize that these steps are implemented {\it without\/}
specifying any boundary conditions for eq.~(2).  The implementation of step~(c)
depends on the particular type of boundary conditions.

We begin by finding a Lax pair.
Let $z=x+iy$.  Then
$\partial_z=(1/2)(\partial_x-i\partial_y)$,
$\partial_\z=(1/2)(\partial_x+i\partial_y)$.
Using these formulas, eq.~(2) can be rewritten in the form
$q_{z\z}-\b^2q=0$.
This equation is the compatibility condition of the following equations:
$$
\mu_{z}+ik\mu=q_z\;,\qquad   
\mu_{z\z}-\b^2\mu=0\;.   \eqno(4)
$$
Indeed, applying the operator $\partial^2_{z\z}-\b^2$ to the first of
eqs.~(4) and using the fact that it commutes with $\partial_z+ik$, we find
$(\partial^2_{z\z}-\b^2)q_z=(\partial_z+ik)(\partial^2_{z\z}-\b^2)\mu=0$.
Differentiating the first of eqs.~(4) and using the remaining eq.~and
$q_{z\z}-\b^2q=0$, it follows that
$$
\mu_{\z}-i{\b^2\over k}\mu=-i{\b^2\over k}q\;.   \eqno(5)
$$
Finally, reverting to the $x$ and $y$ variables the first of eqs.~(4) and
eq.~(5) yield
$$
\mu_x+i(k-\bk)\mu=q_z-i\bk q\;, \qquad
\mu_y-(k+\bk)\mu=iq_z-\bk q\;.  \eqno(6)
$$
{\it In summary\/}, if $q$ satisfies eq.~(2), eqs.~(6) are compatible.

In what follows we will construct a function $\mu(x,y,k)$ which solves both
eqs.~(6) and which is {\it sectionally\/} analytic in the entire complex plane,
including infinity.  The function $\mu$ has different representations in
different sectors of the complex $k$-plane, namely $\mu_0(x,y,k)$, $\kR\geq0$;
$\mu_1(x,y,k)$, $\kR\leq0$ and $\kI\geq-\kR$; $\mu_2(x,y,k)$, $\kR\leq0$ and
$\kI\leq-\kR$, see figure~1.
The functions $\mu_0$, $\mu_1$, $\mu_2$ are defined as follows:
$$
\mu_0(x,y,k)=-\int_y^\infty E_1(y-y',k)[iq_z(x,y')-{\bk}q(x,y')]\,dy'\;, 
\eqno(7)
$$
$$
\mu_j(x,y,k)=E_1(y-x,k)F_j(x,k) +
\int_x^y E_1(y-y',k)[iq_z(x,y')-\bk q(x,y')]\,dy',\quad j=1,2\;, \eqno(8)
$$
where
$$
\eqalign{
&F_1(x,k)=-(1+i)\int_x^\infty E_2(x-x',k)
  [q_z(x',x')-\bk q(x',x')]\,dx'\;, \cr 
&E_1(y,k)=e^{(k+\bk)y}\;,\qquad E_2(x,k)=e^{[(1-i)k+(1+i)\bk]x}\;,}
  \eqno(9)
$$
and $F_2$ is defined by a similar expression with $\int_0^x$ instead of
$-\int_x^\infty$.

In order to derive these formulas we begin with the second of eqs.~(6).  A
particular solution of this equation is given by eq.~(7).  Assuming that
$q(x,y)$,
$q_x(x,y)$ and $q_y(x,y)$ tend to zero as $x\to\infty$, it follows that $\mu_0$
also satisfies the first of eqs.~(6).  We note that
${\rm Re}\{k+\b^2/k\}=\kR(1+{\b^2/|k|^2})$,
thus, since $y-y'<0$, the exponential appearing in~(7) is bounded at infinity
for $\kR\geq0$.  Also ${\rm Re}\{\b^2/k\}={\b^2\kR/|k|^2}$, thus, since
$y-y'<0$, this exponential is bounded at zero for $\kR\geq0$.
The general solution of the second of eqs.~(6) is given by the r.h.s. of
eq.~(8), where
$F_j(x,k)$ is replaced by $F(x,k)\equiv\mu(x,x,k)$.  We choose the function
$F(x,k)$ in such a way that $\mu$ also satisfies the first of eqs.~(6):
$$
{dF\over
dx}={\partial\mu(x,x,k)\over\partial x}+
 {\partial\mu(x,x,k)\over\partial y}=[(1-i)k+(1+i)\bk]F +
 (1+i)[q_z(x,x)-\bk q(x,x)]\;.
\eqno(10)
$$
The integral term of eq.~(8) is bounded for $\kR\leq0$, thus $\mu$ is bounded
for
$\kR\leq0$ provided that there exist solutions of eq.~(10) which are bounded
for
$\kR\leq0$.  A particular solution of eq.~(10) is given by eq.~(9).
We note that the real part of the exponent of $E_2$ involves
$(\kR+\kI)(1+{\b^2/|k|^2})x$, and
${\rm Re}\{(1+i)\b^2/k\}=(\kR+\kI){\b^2/|k|^2}$,
thus, since $x-x'<0$, the exponential appearing in $F_1(x,k)$ is bounded both
at infinity and at zero, provided that $\kR+\kI\geq0$.  Hence $\mu_1(x,y,k)$ is
bounded for $\kR\leq0$ and $\kI\geq-\kR$.  Similar considerations apply to the
function $\mu_2$.

The function $\mu(x,y,k)$ is a sectionally analytic function of $k$ which
satisfies
$\mu=-i{q_z\over k}+\O({1\over k^2})$, $k\to\infty$.
Thus, this function can be reconstructed from knowledge of its ``jumps" across
the contour depicted in figure~1.  If we use the convention that the $\oplus$
region is to the left of the orientation of the contour, these jumps are
$\mu_2-\mu_1$, $\mu_2-\mu_0$, and $\mu_0-\mu_1$.  An important result of the
method of~[1] is that these jumps, denoted by $J(x,y,k)$, have an {\it
explicit\/} $x$ and $y$ dependence.  Indeed, eqs.~(6) imply
$J_x+i(k-\b^2/k)J=0$, $J_y-(k+\b^2/k)J=0$.
Thus, $J(x,y,k)=\exp[(k+\b^2/k)y-i(k-\b^2/k)x]\rho(k)$,
where $\rho(k)$ is a function of $k$ only and can be computed by evaluating $J$
on the boundary of the given domain.  This yields
$$
\eqalign{
&\mu_2-\mu_1=E_3\qa(k)\;,\quad-\kI=\kR\leq0\;; \quad
 E_3(x,y,k)=\exp[(k+\b^2/k)y-i(k-\b^2/k)x]\;;\cr  
&\mu_2-\mu_0=E_3\qb(k)\;,\quad \infty\geq\kI\geq0\;;\qquad
\mu_0-\mu_1=E_3[\qa(k) -\qb(k)]\;,\quad \infty\leq\kI\leq0\;,}
    \eqno(11)
$$
$$
\eqalign{
\qa(k)&=(1+i)\int_0^\infty E_2(-x,k)
  [q_z(x,x)-{\b^2\over k}q(x,x)]\,dx\;, \cr
\qb(k)&=\int_0^\infty E_1(-y,k)
  [iq_z(0,y)-{\b^2\over k}q(0,y)]\,dy\;, 
}  \eqno(12)
$$
and $E_1(x,k)$, $E_2(y,k)$ are defined in~(9).
The function $\mu_0(x,x,k)$ in addition to satisfying eq.~(7) evaluated at
$y=x$, also satisfies eq.~(10), {\it i.e.}, it is given by
$$
\mu_0(x,x,k)=-(1+i)\int_x^\infty E_2(x-x',k)
  [q_z(x',x')-\bk q(x',x')]\,dx'\;, \quad \kR\geq0{\rm\ and\ }\kI\geq-\kR\;.
$$
Evaluating this equation at $x=0$ and using eq.~(7) to compute $\mu(0,0,k)$
we find
$$
\qa(k)=\qb(k)\;,\qquad \kR\geq0{\rm\ and\ }\kI\geq-\kR\;. \eqno(13)
$$
This condition implies that $\mu_1$ provides the analytic continuation of
$\mu_0$, {\it i.e.}, there is no jump between $\mu_0$ and $\mu_1$.

Eqs.~(11), together with the behavior of $\mu$ for large $k$, define a scalar
Riemann-Hilbert (RH) problem~[9,10] whose unique solution is
$$
\mu(x,y,k)={1\over2\pi i}\int_{-i\infty}^0 
 {E_3(x,y,\ell)\qb(\ell)\over\ell-k}\,d\ell +
{1\over2\pi i}\int_0^{e^{3i\pi/4}\infty}
 {E_3(x,y,\ell)\qa(\ell)\over\ell-k}\,d\ell\;,
$$
where the exponential $E_3(x,y,k)$ is defined in eqs.~(11). 
Using eq.~(5) to express $q(x,y)$ in terms of $\mu$ and $\mu_\z$ it follows
that
$$
q(x,y)={1\over2\pi i}\int_{-i\infty}^0 
  e^{(k+{\b^2\over k})y-i(k-{\b^2\over k})x}{\qb(k)\over k}\,dk +
{1\over2\pi i}\int_0^{e^{3i\pi/4}\infty} 
  e^{(k+{\b^2\over k})y-i(k-{\b^2\over k})x}{\qa(k)\over k}\,dk\;. 
\eqno(14)
$$
{\it In summary\/}, suppose that $q(x,y)$ satisfies eq.~(2).  Assume that
$q$ and its derivatives decay to zero at infinity.  Furthermore, assume
that appropriate BCs are prescribed at $x=0$ and at $y=x$
so that there exists a global solution.  This global solution is given by
eq.~(14) where $\qa(k)$ and $\qb(k)$ are defined by eqs.~(12). 
Furthermore, the boundary values of $q$ satisfy the global
relation~(13).

We will now use the above formulation to solve a concrete boundary value
problem.  Let $q(x,y)$ satisfy the BCs
$$
q(x,x) = f(x)\;, \qquad
q_x(0,y)+\alpha q(0,y) = g(y) \;, \eqno(15)
$$
where $\a$ is a given constant and $f(x)$, $g(y)$ are given functions
decaying to zero at large $x$ and large $y$ respectively.
The expression for $\qa(k)$ involves $q_x(x,x)$, $q_y(x,x)$ and $q(x,x)$.
Using the first BC in~(15), it follows that
$q(x,x)=f(x)$, $q_x(x,x)=f'(x)-q_y(x,x)$.  Thus
$\qa(k)$ involves the unknown function $q_y(x,x)$ and a known function
expressed in terms of $f'(x)$ and $f(x)$.  Using integration by parts to
simplify the latter expression we find
$$
\qa(k)=-{1+i\over2}f(0) +
  (k-\bk)\int_0^\infty E_1(-x,k)f(x)\,dx
-i\int_0^\infty E_1(-x,k)q_y(x,x)\,dx\;.  \eqno(16)
$$
Similarly, the expression for $\qb$ involves $q_x(0,y)$, $q_y(0,y)$ and
$q(0,y)$.  Using the second BC in~(15), it follows that $q_x(0,y)=g(y)-\a
q(0,y)$.  Thus $\qb(k)$ involves the unknown functions $q(0,y)$ and
$q_y(0,y)$, and a known function expressed in terms of $g(y)$.  Using
integration by parts to simplify the former expression we find
$$
\qb(k)=-\half f(0)
  +{i\over2}\int_0^\infty E_2(-y,k)g(y)\,dy 
+\half(k-\bk-i\a)\int_0^\infty E_2(-y,k)q(0,y)\,dy\;. 
\eqno(17)
$$
Hence, eq.~(13) becomes
$$
\half(k-\bk-i\a)Q_b(k)+iQ_a(k)=F(k)\;,\qquad 
  \kR\geq0{\rm\ and\ }\kI\geq-\kR\;,  \eqno(18)
$$
where $Q_a(k)$ is the last integral term in eq.~(16), $Q_b(k)$ is the
last integral term in eq.~(17) and $F(k)$ is a known function:
$$
F(k)=-{i\over2}f(0) +
  (k-{\b^2\over k})\int_0^\infty E_1(-x,k)f(x)\,dx
  -{i\over2}\int_0^\infty E_2(-y,k)g(y)\,dy\;.
\eqno(19)
$$
Eq.~(18) is one equation for two unknown functions.  However, using the
analytic properties of $Q_a$ and $Q_b$ it is possible to obtain both these
functions through the solution of a scalar RH problem.
Indeed, we first note that $Q_b(k)$ remains invariant under complex
conjugation; thus, taking the complex conjugate of eq.~(18) we find an equation
involving $Q_a(k)$ and $\overline{Q_a(\k)}$, valid for $|\kI|\leq\kR$,
$\kR\geq0$.  The exponentials of $Q_a(k)$ and of $\overline{Q_a(\k)}$
involve $(1-i)k+(1+i)\bk$ and $(1+i)k+(1-i)\bk$.  Under the
transformation $k\to(1+i)k$ these terms become $2k+\bk$ and $i(2k-\bk)$. 
Thus, using $k\to(1+i)k$ in the equation involving $Q_a(k)$ and
$\overline{Q_a(\k)}$ and then following complex conjugation we  obtain an
equation involving $\int_0^\infty q_y(x,x)\exp[i(2k-\bk)x]\,dx$ and the
complex conjugate of this expression, valid for $\kR\geq0$.  Finally,
letting $k\to-k$ in this equation we find an equation valid for all $\kR$:
$$
\Phi^+(k)-{A(k)\overline{A(-\k)}\over A(-k)\overline{A(\k)}}
  \overline{\Phi^+(\k)}=B(k)\;, \qquad \kI=0\;, \eqno(20)
$$
where
$$
\Phi^+(k)=\int_0^\infty e^{i(2k-\bk)x}q_y(x,x)\,dx, \qquad
  A(k)=k^2-{(1+i)\a\over2}k+{i\b^2\over2}\;,  \eqno(21)
$$
and $B(k)=G(k)$ for $\kR\geq0$, $B(k)=\overline{G(-\k)}$ for $\kR<0$,
with
$$
G(k)=i{\overline{A(-\k)}\over\overline{A(\k)}}[
{A(k)\over A(-k)}\overline{F((1-i)\k)}+
{\overline{A(\k)}\over\overline{A(-\k)}}F((1-i)k)
-\overline{F((1+i)\k)}-F((1+i)k)]\;.
\eqno(22)
$$

Since ${\rm Re}\{i(2k-\b^2/k)\}=-\kI(2+{\b^2/|k|^2})$,
${\rm Re}\{-i\b^2/k\}=-\kI{\b^2/|k|^2}$, $\Phi^+(k)$ is analytic for
$\kI\geq0$, thus $\overline{\Phi^+(\k)}$ is analytic for $\kI\leq0$. 
These facts, together with the estimate $\Phi^+(k)=\O({1\over k})$,
$k\to\infty$, imply that eq.~(20) defines an inhomogeneous scalar RH
problem.  The solution of this problem depends on the {\it index\/} of the
jump function.  Let $\C^+$ ($\C^-$) denote the upper (lower) half of the
complex $k$-plane.  In our case the index is given by $($number of zeros of
$A(k)$ in $\C^+)+($number of zeros of $\overline{A(-\k)}$ in
$\C^+)-($number of zeros of $A(-k)$ in $\C^+)-($number of zeros of 
$\overline{A(\k)}$ in $\C^+)$.  If $N^\pm$ denotes the number of zeros in
$\C^\pm$, the index equals $2N^+-2N^-$.  Thus there exist $3$ cases:
(a)~$A(k)$ has $1$ zero in $\C^+$, then the index is zero and eq.~(20)
has a unique solution. (b)~$A(k)$ has $2$ zeros in $\C^+$, then the index
is $4$ and the solution of~(20) depends on $3$ arbitrary constants.
(c)~$A(k)$ has $2$ zeros in $\C^-$, then the index is $-4$ and the
solution of~(20) exists only if the function $B(k)$ satisfies $4$
orthogonality conditions.
The roots of $A(k)=0$ are given by
$4\lambda=(1+i)\a(1\pm\sqrt{1-({2\b\over\a})^2})$.  Thus if
$|{\b\over\a}|<{\sqrt{2}\over2}$, then $N^+=2$ if $\a>0$ and $N^-=2$ if
$\a<0$.  If $|{\b\over\a}|>{\sqrt{2}\over2}$ then $N^+=1$.  

{\it In summary\/}, if
$|{\b\over\a}|>{\sqrt{2}\over2}$ then eq.~(2) with the BCs~(15) has a unique
solution; if
$|{\b\over\a}|<{\sqrt{2}\over2}$ and $\a>0$, there exist $3$ arbitrary
constants; if $|{\b\over\a}|<{\sqrt{2}\over2}$ and $\a<0$ there exists a
solution only if $f(x)$ and $g(y)$ are such that they satisfy $4$
orthogonality conditions.

We now consider the case of $N^+=N^-=1$.  Let $\lambda^+$ and $\lambda^-$
denote the roots of $A(k)$ in $\C^+$ and $\C^-$ respectively.  Then the
homogeneous version of eq.~(20) becomes
$$
X^+(k)-{(k-\lambda^+)(k-\lambda^-)\,
 (k+\overline{\lambda^+})(k+\overline{\lambda^-})\over
(k+\lambda^+)(k+\lambda^-)\,
 (k-\overline{\lambda^+})(k-\overline{\lambda^-})}X^-(k)=0,
\qquad\kI=0\;.  \eqno(23)
$$
The canonical solution of this problem, {\it i.e.}, the solution
satisfying $X(k)=1+\O({1\over k})$, $k\to\infty$, is given by
$$
X^+(k)={(k-\lambda^-)(k+\overline{\lambda^-})\over
  (k+\lambda^+)(k-\overline{\lambda^+})}\;, \qquad
X^-(k)=\overline{X^+(\k)}\;.  \eqno(24)
$$
Thus,
$$
\Phi(k)={X(k)\over2\pi i}\int_{-\infty}^{\infty}
  {B(\ell)\over X^+(\ell)(\ell-k)}\,d\ell\;.  \eqno(25)
$$

When $N^-=2$, the solution is still given by eq.~(25), but now
$X^+(k)=A(k)\overline{A(-\k)}$. In this case, however, the solution 
exists only if the orthogonality relations 
$$
\int_{-\infty}^{\infty}{B(\ell)\ell^n\over X^+(\ell)}\,d\ell=0\;,
\qquad n=0,1,2,3\;,  \eqno(26)
$$ 
are satisfied.

In the practical physical application of diffusion-limited coalescence, the
BCs~(3) imply
$\a=\gamma'/2-\gamma$,
$f(x)=0$,
$g(y)=(\gamma'-\gamma)e^{-{\gamma\over2}y}$. The three cases for the
zeros of $A(k)$ discussed above correspond to: ($i$)~$\gamma'<0$, for
$N^+=2$, $N^-=0$, ($ii$)~$0\leq\gamma'\leq\gamma$, for
$N^+=N^-=1$, and ($iii$)~$\gamma'>\gamma$, for $N^+=0$, $N^-=2$.  Thus,
case~($i$) is unphysical.  The existence of a unique solution for
case~($ii$) is consistent with the fact that at the endpoints,
$\gamma'=0,\gamma$, solutions were already known~[7,8].  For case~($iii$), it
can be easily verified that the orthogonality conditions~(26) are satisfied,
and hence there exists a unique steady-state solution even when
$\gamma'>\gamma$.  As a final remark, we note that $q_y(x,x)$, and hence the
concentration profile $c(x)=\gamma-q_y(x,x)$, are already derivable from
$\Phi(k)$.  Indeed, the substitution $i(2k-\b^2/k)=-s$ in $\Phi(k)$ yields
the Laplace transform of $q_y(x,x)$, $\int_0^\infty e^{-sx}q_y(x,x)\,dx$,
which is then inverted to find the concentration.

We thank the NSF and the EPRSC for support of this work.

\vfill\eject\NI
{\bf REFERENCES}
\bigskip
\item{[1]} A.~S.~Fokas, ``A unified transform method for solving linear and
certain nonlinear PDEs", {\it Proc. R. Soc. Lond. A} {\bf453}, 1411--1443
(1997).

\item{[2]} A.~S.~Fokas and C.~R.~Menyuk, ``Integrability and self-similarity
in transient stimulated Raman Scattering", {\it J. Nonlinear Sci.} {\bf9},
1--31 (1999).

\item{[3]} A.~S.~Fokas and B.~Pelloni, ``A method of solving moving boundary 
value problems for integrable evolution equations", preprint.

\item{[4]} A.~S.~Fokas and L.~Y.~Sung, ``Initial boundary value problems for
integrable evolution equations on the half-line", preprint.

\item{[5]} A.~S.~Fokas and A.~A.~Kapaev, ``A transform method for the Laplace
equation in a polygon", preprint.

\item{[6]} C. R. Doering, ``Microscopic spatial 
correlations induced by external noise in a reaction-diffusion system",
{\it Physica A} {\bf 188}, 386 (1992).

\item{[7]} D.~ben-Avraham, ``Diffusion-limited coalescence,
$A+A\rightleftharpoons A$, with a trap", {\it Phys. Rev. E} {\bf 58}, 4351
(1998); ``Complete exact solution of diffusion-limited coalescence,
$A+A\to A$", {\it Phys. Rev. Lett.} {\bf 81}, 4756 (1998); ``Inhomogeneous
steady-states of diffusion-limited coalescence,
$A+A\rightleftharpoons A$", {\it Phys. Lett. A} {\bf 249}, 415--423 (1998).

\item{[8]} A.~Donev, J.~Rockwell, and D.~ben-Avraham, ``Generalized von
Smoluchowski model of reaction rates, with reacting particles and a mobile
trap", {\it J. Stat. Phys.} {\bf 95} 97--112 (1999).

\item{[9]} P.~Deift, {\it Orthogonal Polynomials and Random Matrices: A
Riemann-Hilbert Approach}, (Courant Institute Lecture Notes, 1999).

\item{[10]} M.~J.~Ablowitz and A.~S.~Fokas, {\it Complex Variables:
Introduction and Applications\/}, (Cambridge University Press, 1997).

\vfill\eject\NI
{\bf FIGURE CAPTIONS}
\bigskip
\NI{\bf Figure 1:} The regions of analyticity of $\mu_0$, $\mu_1$, $\mu_2$ in
the complex $k$-plane.  The vertical line is $\kR=0$, the other ray is
$\kI=-\kR$ and $\kR<0$.

\bye